\documentclass[twocolumn,secnumarabic,amssymb, nobibnotes, aps, prd,nofootinbib,superscriptaddress]{revtex4-2}

\usepackage[T1]{fontenc} 

\usepackage[T1]{fontenc} 
\usepackage[utf8]{inputenc} 
\usepackage{microtype} 
\usepackage{lmodern} 
\usepackage{booktabs} 
\usepackage{mdframed} 
\usepackage{graphicx}
	
\usepackage[backgroundcolor=white]{todonotes}

\usepackage{comment}
	
\usepackage{hyperref} \usepackage{color}
\definecolor{dark-red}{rgb}{0.4,0.15,0.15}
\definecolor{dark-blue}{rgb}{0.15,0.15,0.4}
\definecolor{medium-blue}{rgb}{0,0,0.5} \hypersetup{colorlinks,
  linkcolor={dark-red}, citecolor={dark-blue}, urlcolor={medium-blue}}

\usepackage{amsmath,amsfonts,amssymb}
\usepackage{eucal} 
\usepackage{mleftright} \mleftright
\usepackage{tensor} 
\usepackage{braket} 
\usepackage{bm} 
\usepackage{mathrsfs} 

\providecommand*{\pd}{\mathop{}\!\partial}
\renewcommand*{\pd}{\mathop{}\!\partial}
\providecommand*{\vd}{\mathop{}\!\delta}
\renewcommand*{\vd}{\mathop{}\!\delta}

\providecommand*{\ct}{{\tilde c}}
\renewcommand*{\ct}{{\tilde c}}
\providecommand*{\ut}{{\tilde u}}
\renewcommand*{\ut}{{\tilde u}}
\newcommand{\bA}{{\overline{A}}}
\newcommand{\bP}{{\overline{P}}}

\renewcommand{\th}{{\theta}}
	
\providecommand*{\xb}{\bm{x}} \renewcommand*{\xb}{\bm{x}}
\providecommand*{\zb}{\bm{z}} \renewcommand*{\zb}{\bm{z}}
\providecommand*{\pb}{\bm{p}} \renewcommand*{\pb}{\bm{p}}
\providecommand*{\db}{\bm{d}} \renewcommand*{\db}{\bm{d}}

\providecommand*{\nb}{\bm{n}} \renewcommand*{\nb}{\bm{n}}
 
\providecommand*{\vb}{\bm{v}} \renewcommand*{\vb}{\bm{v}}

\begin{document}
		
\title{Fracton infrared triangle}
		
\author{Alfredo Pérez}%
\email{alfredo.perez@uss.cl}
\affiliation{Centro de Estudios Científicos (CECs), Avenida Arturo Prat 514, Valdivia, Chile}
\affiliation{Facultad de Ingeniería, Arquitectura y Diseño, Universidad San Sebastián, sede Valdivia, General Lagos 1163, Valdivia 5110693, Chile}
\author{Stefan Prohazka}%
\email{stefan.prohazka@univie.ac.at}
\affiliation{University of Vienna, Faculty of Physics, Mathematical Physics, Boltzmanngasse 5, 1090, Vienna, Austria}
\author{Ali Seraj}%
\email{a.seradg@gmail.com}
\affiliation{Centre de Physique Théorique, Ecole Polytechnique, CNRS, Institut Polytechnique de Paris, 91128 Palaiseau Cedex, France}
		
\begin{abstract}
In theories with conserved dipole moment, isolated charged particles (fractons) are immobile, but dipoles can move. We couple these dipoles to the fracton gauge theory and analyze the universal infrared structure. This uncovers an observable double kick memory effect which we relate to a novel dipole soft theorem. Together with their asymptotic symmetries this constitutes the first realization of an infrared triangle beyond Lorentz symmetry. This demonstrates the robustness of these IR structures and paves the way for their investigation in condensed matter systems and beyond.
\end{abstract}
		
\maketitle
	
\section{Introduction}
\label{sec:introduction}
		
Fractons~\cite{Chamon:2004lew,Haah:2011drr} are novel quasiparticles
whose characteristic feature is their limited mobility. This
restricted mobility originates from their built-in \textit{dipole
  symmetry} which leads to conserved dipole moments
$d^{i}=\int x^{i}\rho d^{3}x$. Fracton theories attract attention not
only for their interesting phenomenological applications, but also for
their intricate theoretical underpinnings which challenge common
quantum field
techniques~\cite{Nandkishore:2018sel,Pretko:2020cko,Grosvenor:2021hkn}.
The dipole symmetry is an example of generalized symmetries, whose
study has led to breakthroughs in our understanding of quantum field
theories~\cite{Brauner:2022rvf,Cordova:2022ruw}. Moreover, the
underlying symmetries are closely related to Carroll
symmetries~\cite{Bidussi:2021nmp,Marsot:2022imf,Figueroa-OFarrill:2023vbj,Figueroa-OFarrill:2023qty},
which play a fundamental role in flat-space holography
\cite{Figueroa-OFarrill:2021sxz,Donnay:2022aba,Bagchi:2022emh,Donnay:2022wvx}.
		
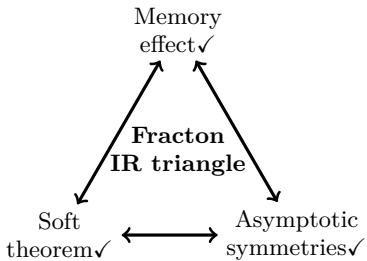
\begin{figure}
\centering
\begin{tikzpicture}[scale=0.35]
  \node[align=center] (st) at (-4.5,0) {Soft\\theorem$\checkmark$};
  \node[align=center] (as) at (4.5,0)
  {Asymptotic\\symmetries$\checkmark$}; \node[align=center] (me) at  (0,7.8) {Memory\\effect$\checkmark$};
  \node[align=center] (fr) at(0,3.2) {\textbf{Fracton}\\\textbf{IR triangle}};

\draw[very thick,<->] (st) -- (as);
\draw[very thick,<->] (st) -- (me);
\draw[very thick,<->] (as) -- (me);
\end{tikzpicture}
\caption{Fracton infrared triangle}
\label{fig:frac-triangle}
\end{figure}

The infrared (IR) triangle~\cite{Strominger:2017zoo} is a triangular
correspondence, which connects asymptotic symmetries, memory effects
and soft theorems. It controls the infrared behavior of gravity and
several relativistic gauge theories, and is a building block of the
celestial holography
program~\cite{Raclariu:2021zjz,Pasterski:2021rjz,Pasterski:2021raf,McLoughlin:2022ljp,Donnay:2023mrd}.
In this work, we establish the infrared triangle in fracton gauge
theory by demonstrating the existence of a novel and observable double
kick memory effect, which we also relate to a dipole soft theorem and
novel asymptotic symmetries (see Figure~\ref{fig:frac-triangle}). This
shows that IR triangles indeed persist beyond the Lorentzian setup and
lead to exciting new physics.

One consequence of the non-lorentzian nature of the fracton gauge
theory is that its degrees of freedom have two different dispersion
relations and propagation speeds. As a result, the spacetime has two
radiative regions (see Figure~\ref{fig:double-kick}), in
contradistinction to the single null infinity in lorentzian theories.
This makes the asymptotic structure of these theories richer than that
of their relativistic counterparts. These novel features also imply
the need for implementing new techniques beyond the conventional ones.
For example, because of the asymmetry between space and time, a
radiative coordinate system of Bondi type is not directly available.
		
One of our main results is a novel memory effect. Memory effects refer
to observables that persist in a probe system after the passage of
waves. Early examples of memory effects in the context of gravity
include the displacement of freely falling test masses
\cite{Zeldovich:1974gvh,Braginsky:1985vlg,braginsky1987gravitational,Christodoulou:1991cr}.
However, this field has witnessed a significant interest in recent
years due to discovery of the relation between memory effects and
fundamental properties of gravity
\cite{Strominger:2013lka,Strominger:2014pwa,Strominger:2015bla}, and a
plethora of novel persistent wave observables have been found in
gravity and gauge theory
\cite{Bieri:2013ada,Bieri:2013hqa,Pasterski:2015tva,Pasterski:2015zua,Strominger:2017zoo,Flanagan:2019ezo,Nichols:2018qac,Seraj:2021qja,Seraj:2021rxd,Seraj:2022qyt,Seraj:2022qqj,Oblak:2023axy}.
While fracton memory effect shares some features with gravity and some
with gauge theories, it has unique properties, in particular, it leads
to a \textit{double kick} effect on test quasi-particles (dipoles), as
is depicted in Figure~\ref{fig:double-kick}. This double kick memory
effect is a measurable infrared observable and encourages the exciting
perspective to try to observe them in condensed matter systems,
especially considering the ongoing experimental investigations into
systems with dipole symmetry
\cite{Guardado-Sanchez:2019bjm,Scherg:2020mcp,2021arXiv210615586K,2022PhRvX..12b1014Z}.
		
This work is structured as follows. In Section~\ref{sec:setup} we
review the fracton scalar charge gauge theory. In
Section~\ref{sec:Radiation} we derive the spectrum of the theory and
decoupled equations obeyed by each sector. In particular, we obtain
the analogue of the Liénard-Wiechert potentials, and thereby uncover
the memory effects in the scattering of dipoles. We show that test
dipoles experience a double kick memory effect. In
Section~\ref{sec:Asymptotic} we analyze asymptotic symmetries and
their corresponding conservation laws, which we relate in
Section~\ref{sec:Memory} to memory effects. In
Section~\ref{sec:soft-factor-from} we derive the dipole soft factor
from the memory effect (an alternative derivation using Feynman
diagrams is presented in the supplementary
material~\ref{sec:soft-factor-feynman}). We finish with a discussion
and an outlook in Section~\ref{sec:discussion}. Further details will
appear in our extended work~\cite{Perez:2024xxx}.
		
\section{Fracton gauge theory}
\label{sec:setup}
		
In this section, we introduce the (scalar charge) gauge
theory~\cite{Pretko:2016kxt,Pretko:2016lgv} which describes the
interactions among charged fractons. It is a higher-rank gauge theory
defined by a scalar field $\phi$ and a symmetric tensor $A_{ij}$
($i,j,k$ are spatial indices from $1$ to $3$), with the Lagrangian
density
\begin{align}
  \label{eq:lagran}
  \mathcal{L}[A_{ij},\phi] &= \frac{1}{2} E_{ij} E\indices{^{ij}} - \frac{c^{2}}{4} F_{ijk}F^{ijk} +  \phi \rho  - A_{ij} J^{ij},
\end{align}
where
\begin{align}
  E_{ij}&= \pd_{t}A_{ij} - \pd_{i}\pd_{j}\phi &  F_{ijk}&= 2 \pd_{[i}A_{j]k} \, .
\end{align}
The constant $c$ has units of velocity, but is not necessarily the
speed of light. The tensors $E_{ij}$ and $F_{ijk}$ are analogous to an
electric and magnetic tensor, respectively. The symmetries imply
$F_{[ijk]}=0$ and the useful relation
$F_{i[jk]} = -\frac{1}{2} F_{jki}$. The action \eqref{eq:lagran} is
invariant under the gauge transformation
\begin{align}
  \label{eq:gauge-transf}
  \vd_{\Lambda} \phi &= \pd_{t}\Lambda  & \vd_{\Lambda}A_{ij} = \pd_{i}\pd_{j} \Lambda \, ,
\end{align}
and leads to the equations of motion
\begin{subequations}
  \label{eq:EOM}
  \begin{align}
    \partial_i \partial_j E\indices{^{ij}}  &= \rho \label{eq:EOMScalargauge1}
    \\
    \pd_{t} E_{ij} - c^{2} \pd_{k} F\indices{^{k}_{(ij)}} &= - J_{ij} \, , \label{eq:EOMScalargauge2}
  \end{align}
\end{subequations}
where $\rho$ and $J_{ij}$ represent the charge and current densities,
respectively. Consistency of the field equations~\eqref{eq:EOM} leads
to the continuity equation
\begin{align}
  \label{eq:continuity}
  \pd_{t} \rho + \pd_{i}\pd_{j} J^{ij} = 0 \, 
\end{align}
which implies that the electric charge $Q$ and dipole moment $d^{i}$
\begin{align}
  \label{eq:charge-dipolemom}
  Q&=\int \rho \, d^3x  & d^i&=\int x^i \rho  \, d^3x 
\end{align}
of a localized source are conserved. The conservation of the dipole
moment implies, in particular, that isolated monopoles in this theory
cannot move.
		
\section{Decoupled field equations and memory effect}
\label{sec:Radiation}
		
An important consequence of the fact that the fracton gauge theory is
not Lorentz invariant is that various dynamical degrees of freedom
obey different dynamical equations. We decouple these equations of
motion using a systematic decomposition of the gauge field $A_{ij}$
into representations of the rotation group. Moreover, imposing the
constraint \eqref{eq:EOMScalargauge1}, one finds
\begin{align}\label{eq:Edecomposed}
	E_{ij}&=\dot A_{ij}^{\mathrm{T}}+\dot A_{ij}^{\mathrm{TL}}+\pd_i\pd_j\Delta^{-1}(\Delta^{-1}\rho)\,,
\end{align}
where the superscripts T and TL, denoted collectively by $\#$, refer
to transverse and transverse-longitudinal projections
$A^\#_{ij}=P^\#_{ijmn}A_{mn}$ with
\begin{align}
	P^\mathrm{T}_{ijmn}&=P_{m(i}P_{j)n} &  P^\mathrm{TL}_{ijmn}&=2P_{m(i}\Pi_{j)n}\,,
\end{align}
defined in terms of transverse and longitudinal projectors
\begin{align}
	P_{ij}&=\delta_{ij}-\Pi_{ij} & \Pi_{ij}&=\pd_i\pd_j \Delta^{-1}
\end{align}
where $\Delta^{-1}$ is the inverse of the Laplacian
$\Delta=\pd_i\pd^i$, given by a Green function integral. Using
\eqref{eq:Edecomposed} in the dynamical equation
\eqref{eq:EOMScalargauge2} combined with
$\dot F_{ijk}= 2 \pd_{[i}E_{j]k}$, one finds the decoupled dynamical
equations
\begin{align}
	\label{eq: wave eqns}
	\Box_{c} \,A_{ij}^{\mathrm{T}} &= J_{ij}^{\mathrm{T}} &
	\Box_{\ct} \,A_{ij}^{\mathrm{TL}} &=  J_{ij}^{\mathrm{TL}} 
\end{align}
where $\Box_{c}\equiv -\pd_t^2+c^2\Delta$ is the wave operator with
speed $c$. The unequal propagation speeds $c$ and $\ct=c/\sqrt{2}$ of
the dynamical degrees of freedom reflect the non-lorentzian nature of
this theory, whose Aristotelian symmetry structure is discussed
in~\cite{Bidussi:2021nmp,Jain:2021ibh}. Also note that gauge
transformations \eqref{eq:gauge-transf} whose parameter vanishes at
the boundary (i.e., when $\Lambda=O(1/r)$), are in the kernel of
$P^{\mathrm{T}}$ and $P^{\mathrm{TL}}$. This implies that the
dynamical variables $A_{ij}^{\mathrm{T}}, A_{ij}^{\mathrm{TL}}$ are
gauge invariant and account for the expected $2+1$ and $2$ degrees of
freedom, respectively. However, they do transform under so-called
``large-gauge transformations'' as will be discussed in section
\ref{sec:Asymptotic}. Equations~\eqref{eq: wave eqns} are solved by
\begin{align}
	\label{eq:Green solution}
	A_{ij}^\mathrm{T}&=P^\mathrm{T}_{ijmn}\Box_{c}^{-1} J_{mn} &  A_{ij}^{\mathrm{TL}}&=P^{\mathrm{TL}}_{ijmn}\Box_{\ct}^{-1} J_{mn}
\end{align}
where $\Box_{c}^{-1}$ represents the inverse of 
$\Box_c$ using a retarded Green function integral. In deriving \eqref{eq:Green solution}, we
have used the commutativity of $\Box_{c}^{-1}$ and the projectors.

\paragraph*{Asymptotic behavior.}

Assuming that the source is localized, the asymptotic behavior of the
fields can be derived from an asymptotic expansion of \eqref{eq:Green
	solution} in the radiation zone $r\to \infty$. One finds that
$A_{ij}=\frac{1}{r}\overline{A}_{ij}+O(1/r^2)$. Moreover, the
Coulombic contribution to \eqref{eq:Edecomposed} takes the asymptotic
form 
\begin{align}
	\Delta^{-1}(\Delta^{-1}\rho)&=\frac{1}{8\pi}\big(-Q\,r+d_i n^i+O(1/r)\big)
\end{align}
and therefore
\begin{align}\label{eq:electric field asymptotics}
	E_{ij}&=\frac{1}{r}\left[\dot \bA{}_{ij}^{\mathrm{T}}+ \dot \bA{}_{ij}^{\mathrm{TL}}-\frac{Q}{8\pi}\overline{P}_{ij}\right]+O(1/r^2)\,,
\end{align}
where $\bP_{ij}=\delta_{ij}-n_in_j$ and
$n^i= \allowbreak (\sin\th\cos\varphi, \allowbreak
\sin\th\sin\varphi,  \allowbreak\cos\th)$ is the radial normal vector in spherical
coordinates $\th^A=(\th,\varphi)$.

\subsection{Radiation of scattering dipoles}
\label{sec:moving-dipoles}

Since in this theory isolated monopoles cannot move, a natural setup
is the scattering of moving dipoles. The current of a dipole $d^{i}$
moving on a path ${z}^{i}(t)$ with velocity $v^{i} = \dot z^{i}(t)$ is
given by~\cite{Pretko:2016lgv}
\begin{subequations}
	\begin{align}
		\label{eq:dipole-current}
		\rho(t,\xb) &= -d^{i} \pd_{i}\delta^{3} (\xb-\zb(t)) \\ 
		J_{ij}(t,\xb) & =- d_{(i}v_{j)}\delta^{3} (\xb-\zb(t)) \, .
	\end{align}
\end{subequations}
We can use the current to calculate the Green integral
\begin{align}
	\label{eq:lien}
	\Box_{c}^{-1} J_{ij} = \frac{1}{4\pi c^{2}}
	\left.
	\frac{d_{(i}v_{j)}}{R (1-\hat{\bm{R}} \cdot \bm{v}/c)}
	\right|_{t_\text{ret}=t-R/c} \, ,
\end{align}
where $\bm{R}\equiv \bm{x}-\bm{z}(t)$ and $R$, $\hat{\bm{R}}$ are its
norm and unit direction, respectively. This is the fracton dipole analog of
the Liénard-Wiechert solution of moving point charges in
electrodynamics. Inserting \eqref{eq:lien} into \eqref{eq:Green
	solution}, one finds that far from the source, asymptotic fields
$\bA{}^\mathrm{T}_{ij}(\nb)$ and
$\bA{}^{\mathrm{TL}}_{ij}(\nb)=n_{(i}\bA{}^{\mathrm{TL}}_{j)}$ of a dipole
moving with constant velocity are given by
\begin{align}
	\hspace{-.2cm} \bA{}^\mathrm{T}_{ij}= \frac{1}{4\pi c^{2}}
	\frac{d^\perp_{(i}v^\perp_{j)}}{ 1- \nb \cdot \bm{v}/c}, \quad
	\bA{}^{\mathrm{TL}}_{i}\!\!=\frac{1}{4\pi \ct^{2}}
	\frac{d^\perp_{i}v_{r}+v^\perp_{i}d_{r}}{ 1- \nb\cdot \bm{v} /\ct}
\end{align}
where we use $X^\perp_i\equiv \overline{P}_{ij}X^j$ and
$X_r\equiv X_i n^i$.

\paragraph*{Memory effects.}

The scattering process of $N$ dipoles labeled by
$\alpha=\{1,\cdots ,N\}$ will induce a memory
$\delta\bA{}^\mathrm{T}_{ij}=\lim_{u\to\infty}(\bA{}^\mathrm{T}_{ij}(u)-\bA{}^\mathrm{T}_{ij}(-u))$,
$\delta\bA{}^{\mathrm{TL}}_{ij}=\lim_{\ut\to\infty}(\bA{}^{\mathrm{TL}}_{ij}(\ut)-\bA{}^{\mathrm{TL}}_{ij}(-\ut))$
given by
\begin{subequations}
	\label{eq:memory_dip}
	\begin{align}
		\delta\bA{}^\mathrm{T}_{ij}(\bm{n})&= \frac{1}{4\pi c^{2}} \sum_{\alpha=1}^N
		\frac{\eta^\alpha \, d^{\alpha\perp}_{(i}\,v^{\alpha\perp}_{j)}}{ 1-\bm{n} \cdot \bm{v}^\alpha /c} \label{memory T} \\
		\delta\bA{}^{\mathrm{TL}}_{i}(\bm{n})&=\frac{1}{4\pi \ct^{2}}\sum_{\alpha=1}^N
		\frac{\eta^\alpha\,(d^{\alpha\perp}_{i}v^\alpha_{r}+v^{\alpha\perp}_{i}d^{\alpha}_r)}{ 1-\bm{n} \cdot \bm{v}^\alpha/\ct} \, , \label{memory TL}
	\end{align}
\end{subequations}
where $\eta=1$ for outgoing and $-1$ for incoming dipoles.

\paragraph*{Double kick memory effect.}

\begin{figure}
	\centering
	\includegraphics[width=8cm]{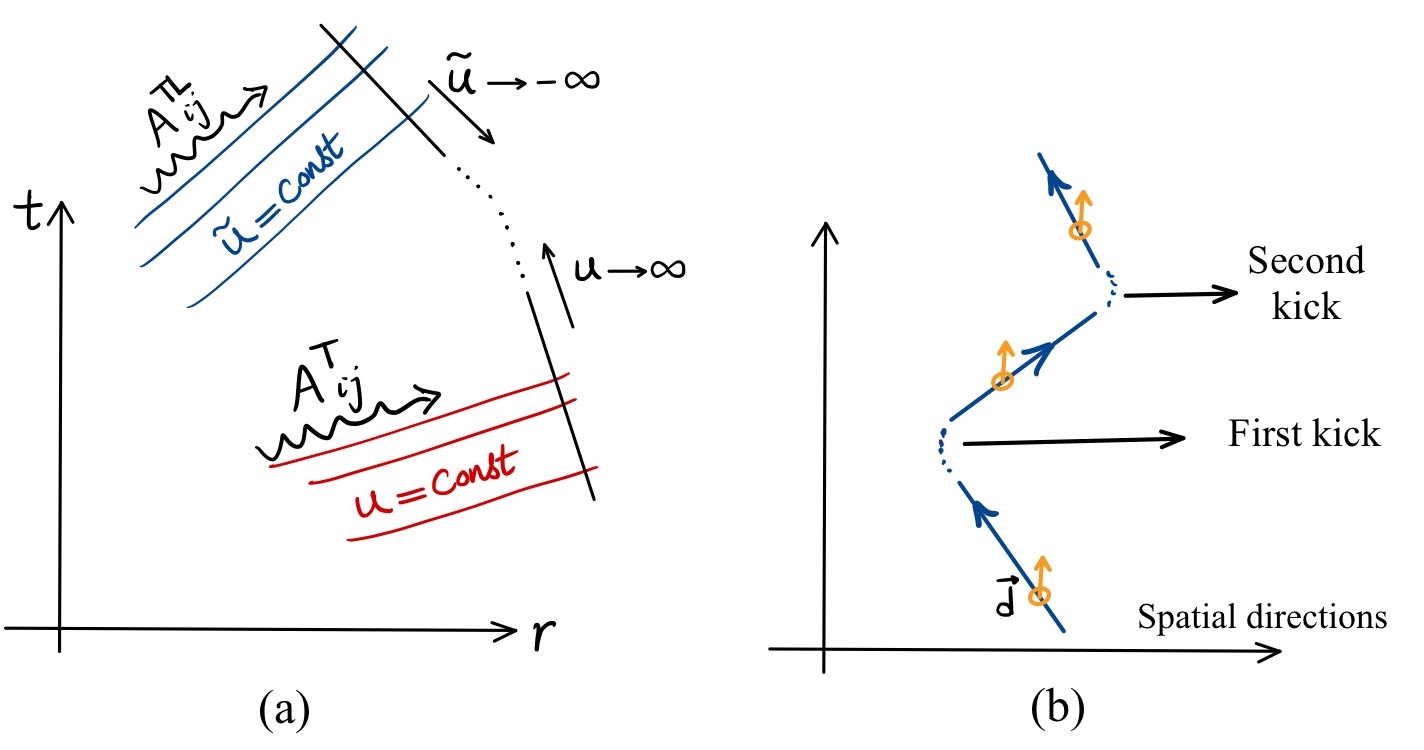} 
	\caption{(a): Since the waves propagate with speeds $c$ and $\ct$
		the theory has two different radiation zones (two ``null
		infinities''). (b): It follows that a dipole in the far region
		will receive two kicks, but the orientation $\vec d$ of the dipole
		stays inert.}
	\label{fig:double-kick}
\end{figure}

These memory effects have observable consequences. As an example,
consider a fractonic dipole situated at far distance from the
source of radiation. The dipole is affected by the radiation through
the generalized Lorentz force law~\cite{Pretko:2016lgv}
\begin{equation}
	\label{eq:Lorentz-force}
	{\dot v}_{i}=-\frac{d^{j}}{m}\left(E_{ij}+v^{k}F_{kij}\right) \, .
\end{equation}
Using the asymptotic form \eqref{eq:electric field asymptotics} of the
electric field and an analogous expression for the magnetic field, and
integrating the result over time in the interval $(t_0,t_f)$, we find
that there is a net kick effect on the dipole that is proportional to
the memory effects
\begin{subequations}
	\label{eq:kick_memory}
	\begin{align}
		\delta v_r&=-\tfrac{d^i}{mr}\left( \delta \bA{}^{\mathrm{TL}}_i - \tfrac{v_0^j}{c} \delta \bA{}^{\mathrm{T}}_{ij}\right)+O(r^{-2}) \\
		\delta v^\perp_i&=\tfrac{-1}{mr}\left[ d_r \,\delta \bA{}^{\mathrm{TL}}_i \!\!+ d^j \delta \bA{}^{\mathrm{T}}_{ij}(1+\tfrac{v_0^r}{c})-\tfrac{q d^\perp_i}{8\pi}\delta u\right]+O(r^{-2}).
		\label{eq:kick_memory-2}
	\end{align}
\end{subequations}
Each of the fast and slow radiative modes cause a net kick effect on
the test dipole and thus it undergoes a \textit{double kick} memory
effect (see Figure~\ref{fig:double-kick}). The term proportional to
$\delta u$ in~\eqref{eq:kick_memory-2} describes the acceleration
produced by the Coulomb field of the total charge of the source, which
in contrast to electromagnetism, is of the same order in the large $r$
expansion as the wave solution.

\section{Asymptotic conditions and Bondi analysis of fractonic waves}
\label{sec:Asymptotic}

An immediate consequence of the different propagation speeds in this
theory is that at very large distances, there will be a decoupling of
the $\mathrm{T}$ and $\mathrm{TL}$ sectors, defining two distinct
radiation zones: the fast radiation zone of $\mathrm{T}$-waves, where
$t,r\rightarrow\infty$ while $u:=t-r/c$ remains finite, and the slow
radiation zone of $\mathrm{TL}$-waves, with $\ut:=t-r/\ct$ finite.
Accordingly, the radiative phase space splits into two distinct phase
spaces $\Gamma_{\rm T}, \Gamma_{\rm TL}$. The splitting is consistent,
since the asymptotic observables we are interested in, such as memory
effects, can be decomposed into operators with support in either of
the phase spaces $O=O_{\rm T}+O_{\rm TL}$. A more comprehensive
analysis which might be useful to study subleading radiative effects,
which are not necessarily separable, would involve considering the
direct product of the two phase spaces. Such analysis requires a
careful analysis of the matching between the fields in the two
radiative zones and goes beyond the scope of this letter.

We will therefore analyze the structure of fields and asymptotic
symmetries in each of these asymptotic regions independently. To this
end, we solve~\eqref{eq: wave eqns} asymptotically in the limit
$r\to\infty$. We assume that the source fields decay fast enough, so
that we can implement source-free wave equations at leading orders. We
use the notation $(c_{\#},u_{\#})$ to unify results that are valid in
both radiative regions with their respective propagation speed and
retarded time. Also, overdot refers to derivative with respect to
retarded time of the region under consideration.
		
The transformation from Cartesian to spherical coordinates, which are
more convenient for the asymptotic analysis is carried out by suitable
projections with the triad $(n^i,re_A{}^i)$ and
$ e_A{}^i(\nb)= \frac{\pd n^i}{\pd \theta^A}$. The induced metric
$\gamma_{AB}=\delta_{ij}e_A{}^ie_B{}^j$ denotes the metric of the unit
2-sphere
$\gamma_{AB}dx^{A}dx^{B}=d\theta^{2}+\sin^{2}\theta d\varphi^{2}$
which is used to lower and raise $A,B,\ldots$ indices and has
determinant $\gamma$. Therefore, the analysis in the preceding
sections reveals the following asymptotic behavior for the electric
field written as a tensor density, i.e., multiplied by
$r^2 \sqrt{\gamma}$
\begin{subequations}
\label{eq:AsympETT}
\begin{align}
E^{rr} & =\overline{E}^{rr}+r^{-1}E_{\left(-1\right)}^{rr}+O\left(r^{-2}\right) \\
E^{rA} & =\overline{E}^{rA}+r^{-1}E_{\left(-1\right)}^{rA}+O\left(r^{-2}\right) \\
E^{AB} & =r^{-1}\overline{E}^{AB}+r^{-2}E_{\left(-2\right)}^{AB}+O\left(r^{-3}\right) \, .
\end{align}
\end{subequations}
The asymptotic behavior of the electric field is consistent with the following fall-off for the potentials
\begin{subequations}
\label{eq:AsympATT}
\begin{align}
    A_{rr} & =O\left(r^{-1}\right) \\
    A_{rA} & =\overline{A}_{rA}+O\left(r^{-1}\right) \\
    A_{AB} & =r\overline{A}_{AB}+O\left(r^{0}\right) \\
    \phi & =\frac{q}{8\pi}r+\phi^{(0)}+O\left(r^{-1}\right) \, .
\end{align}
\end{subequations}
Here, $q$ is a constant parameter of the asymptotic solution, which
after matching with the bulk solution coincides with the total charge
$Q$. As shown in~\cite{Perez:2022kax}, this expression for the leading
order of $\phi$ is essential for achieving both finite energy density
and charges. Note that the fall-off is preserved under the
Aristotelian symmetries, i.e., spacetime translations and rotations.

The fall-off \eqref{eq:AsympATT} is preserved under gauge
transformations of the form
\begin{align}
  \label{asymptotic symmetries}
  \Lambda=r\lambda\left(\theta,\varphi\right)+u\,c\,\eta\left(\theta,\varphi\right)+\epsilon\left(\theta,\varphi\right)+O\left(r^{-1}\right) \,,
\end{align}
and a similar expression in the slow radiation zone by replacing
$(u,c)$ by $(\ut,\ct)$ and $(\lambda,\eta)$ by
$({\tilde\lambda},\tilde\eta)$. The corresponding gauge transformation
take the same functional form in both regions (except that
$(\eta,\lambda)\to({\tilde\lambda},\tilde\eta)$ in the TL sector)
\begin{subequations}
	\label{eq:deltaATTall}
	\begin{align}
		\delta_{\Lambda} \overline{A}_{rA}&=-D_{A}\eta \,,\\
		\delta_{\Lambda} \overline{A}_{AB}&=D_{AB}\lambda+\gamma_{AB}\Big(\frac{1}{2}\left(D^{2}+2\right)\lambda-\eta\Big) \, ,\label{eq:DeltaATT}
	\end{align}
\end{subequations}
where $D_{A}$ is the covariant derivative with respect to
$\gamma_{AB}$, while $D^2=D_AD^A$ is the sphere Laplacian and
$D_{AB}\equiv
\frac{1}{2}\left(D_{A}D_{B}+D_{B}D_{A}-\gamma_{AB}D^{2}\right)$. The
corresponding charges and fluxes should be worked out in each
radiative region separately, as the radiative fields behave
differently:

\paragraph*{Transverse sector $\mathrm{T}$.}

The radiative field in this region is $\overline{A}_{AB}(u,\th^A)$,
while $\overline{A}_{rA}$ and $\phi^{(0)}$ are time independent
functions on the sphere, we therefore find
\begin{subequations}
\label{eq:asy_TT}
\begin{align}
    \overline{E}{}^{rA}&=0& 
    \overline{E}{}^{AB}&=-\sqrt{\gamma}\gamma^{AB}\frac{q}{8\pi}+\sqrt{\gamma}\dot{\overline{A}}^{AB} . 
\end{align}
\end{subequations}
		
\paragraph*{Transverse-longitudinal sector  $\mathrm{TL}$.}

The radiative field in this region is $\overline{A}_{rA}(\ut,\th^A)$,
while $\overline{A}_{AB}$ and $\phi^{(0)}$ are time independent.
Accordingly,
\begin{align}
\label{eq:asy_TL}
    \overline{E}{}^{rA}&=\sqrt{\gamma}\gamma^{AB}\dot{\overline{A}}_{rB} &
    \overline{E}{}^{AB}&=-\sqrt{\gamma}\gamma^{AB}\frac{q}{8\pi} \, .
\end{align}		
The charges corresponding to asymptotic symmetries \eqref{asymptotic
  symmetries} can be worked out using canonical or covariant phase
space methods \cite{Regge:1974zd,Lee:1990nz}. To get finite charges, we
also impose the following constraints~\cite{Perez:2022kax}, consistent
with the solutions obtained in the previous section
\begin{equation}
  \label{eq:conds}  
  \overline{A},\dot{A}^{\left(0\right)},\phi^{\left(0\right)},\lambda,\,\eta\; \text{contain}\;\ell\geq1 \;\text{harmonics}\, ,
\end{equation}
in a harmonic expansion in $Y_{\ell,m}(\theta,\phi)$.
Using the equation of motion 
\begin{equation}
    \frac{1}{c_{\#}}\dot{\overline{E}}{}^{rr}+ \overline{E}-2 D_{A}\overline{E}{}^{rA}=-\sqrt{\gamma}\frac{q}{4\pi} \, ,
\end{equation}
together with the conditions \eqref{eq:asy_TT} or \eqref{eq:asy_TL},
one finds that the total charge is finite and reads
\begin{equation}
    Q\left[\epsilon,\lambda,\eta\right]=\oint d^{2}x\left(-\epsilon\sqrt{\gamma}\frac{q}{4\pi}+\lambda\mathcal{P}+\eta\mathcal{Q}\right),\label{eq:QT}
\end{equation}
where the charge densities depending on the radiation zone are given
by
\begin{subequations}
\begin{align}
    \mathcal{P} & :=\overline{E}{}^{rr}-2D_{A}E_{\left(-1\right)}^{rA}+E_{\left(-2\right)}+\frac{1}{c_{\#}}\dot{E}_{\left(-1\right)}^{rr} \label{eq:PTT}\\
    \mathcal{Q} & :=- \overline{E}{}^{rr} \, .\label{eq:QTT}
\end{align}
\end{subequations}
The term proportional to $\epsilon$ in \eqref{eq:QT} gives the total
charge, while terms proportional to $\lambda$ and $\eta$ give two
infinite sets of charges in each sector, where the conserved dipole
moment is the $\ell=1$ in the mode expansion of $\mathcal{P}$.
		
In the presence of radiation, charges are no longer conserved, but
carried away by fractonic waves. The time evolution of the charges are
specified by the following flux equations derived from the equations
of motion. In the fast radiation zone ($\mathrm{T}$-sector)
\begin{align}
    \dot{\mathcal{P}} & =-\sqrt{\gamma}\left(D^{A}D^{B}+\gamma^{AB}\right)\dot{\overline{A}}_{AB}&
    \dot{\mathcal{Q}} &=c\sqrt{\gamma}\dot{\overline{A}}\,,		\label{eq:fluxTT}
\end{align}
whereas in the slow radiation zone
($\mathrm{TL}$-sector)

\begin{align}
\label{eq:FluxTL}
    \dot{\mathcal{P}}&=0 &\dot{\mathcal{Q}}&=-2\ct\sqrt{\gamma}D^{A}\dot{\overline{A}}_{rA} \, . 
\end{align}
Radiation also carries away energy from the system. The field
equations imply that the energy density
$\mathcal{H}=\frac{1}{2} E_{ij} E\indices{^{ij}} +
\frac{c^{2}}{4}F_{ijk}F^{ijk}$ obeys the following conservation
equation
\begin{equation}
	\partial_{t}\mathcal{H}=\partial^{k}\left(c^{2}E^{ij}F_{kji}\right)\,,\label{eq:fluxH}
\end{equation}
which reveals the following flux-balance equations
\begin{align}
	\frac{dE}{dt}&=-c\oint d^{2}x\,\sqrt{\gamma}\dot{\overline{A}}{}^{AB}\dot{\overline{A}}_{AB}&(\mathrm{T}\text{-sector})\,,\\
	\label{eq:FluxETL}  
	\frac{dE}{dt}&=-2\ct\oint d^{2}x\,\sqrt{\gamma}\gamma^{AB}\dot{\overline{A}}_{rA}\dot{\overline{A}}_{rB}&(\mathrm{TL}\text{-sector})\,.
\end{align}
These are the fractonic analogue of Bondi's energy loss formula
\cite{Bondi:1960jsa,Bondi:1962px}, showing that the energy is always
decreasing in time.
		
\section{Memory effect and asymptotic symmetries}
\label{sec:Memory}
		
In this section, we show that fracton memory effects can be realized
as a vacuum transition under fracton asymptotic symmetries. Consider a
dynamical process in which the system is non-radiative before some
initial time and after some final time, implying that
$\dot{\overline{A}}_{AB}=0$ in the limit $u\to\pm\infty$ and
$\dot{\overline{A}}_{rA}=0$ in the limit $\ut\to\pm\infty$. Therefore,
the initial and final radiative vacua are labeled by time-independent
tensors $\bA_{rA}(\th^A), \bA_{AB}(\th^A)$. The memory effect
discussed in section \ref{sec:moving-dipoles} implies that the vacua
are not identical before and after radiation. Rather, the transition
between the vacua induces a large gauge transformation as we will see
shortly.
				
A generic vector $Y_A$ and symmetric tensor $X_{AB}$ on the sphere can
be decomposed as
\begin{subequations}
  \begin{align}
    Y_A&=D_AY^++\epsilon_A{}^BD_BY^-\,\\
    X_{AB}&=D_{AB} X^++\epsilon_A{}^CD_{BC}X^-+\tfrac{1}{2}\gamma_{AB}X\,,
  \end{align}
\end{subequations}
where the terms involving $\epsilon_{AB}$ tensor have odd (magnetic)
parity. However, starting from \eqref{eq:memory_dip} and transforming
to spherical coordinates, we find that the magnetic terms vanish and
the memory terms can be expressed in terms of three scalar
\textit{memory fields} $C_\mathrm{S},C_\mathrm{V},C_\mathrm{T}$ on the
sphere\footnote{Subscripts $\mathrm{S},\mathrm{V},\mathrm{T}$ refer to
  scalar, vector, and tensor which refer to the corresponding
  propagating modes.}
\begin{align}
  \label{eq:DeltaA_Gold}
    \delta \overline{A}_{AB}&=D_{AB}C_\mathrm{T}+\tfrac{1}{2}\gamma_{AB}C_\mathrm{S}&\delta \overline{A}_{rA}&=0, &(\mathrm{T}\text{-sector})\,,\nonumber\\
\delta \overline{A}_{rA}&=D_{A}C_\mathrm{V} & \delta \overline{A}_{AB}&=0&(\mathrm{TL}\text{-sector})\,.
\end{align}
The first (second) line refers to the memory accumulated during the
fast (slow) radiation.

\paragraph*{Memory as vacuum transition}

In each sector, the corresponding memory leads to a change of vacuum
given by a large gauge symmetry. The memory $\delta \overline{A}_{rA}$
in the TL sector corresponds to a large gauge transformation given by
$\tilde\eta=-C_\mathrm{V},\, \tilde\lambda=0$ in
\eqref{eq:deltaATTall}. The T sector is subtle, the memory
$\delta \overline{A}_{AB}$ corresponds to choosing
$\lambda=C_{\rm T},\, \eta=\tfrac{1}{2}((D^2+2)C_{\rm T}-C_{\rm S})$
in \eqref{eq:deltaATTall}. This choice also induces a change in
$\overline{A}_{rA}$, but that is no problem since $\overline{A}_{rA}$
is not part of the radiative phase space of the T
sector.\footnote{Upon considering the product phase space of the slow
  and fast radiation modes, there could emerge a new symmetry, which
  is exclusively responsible for the memory in the trace mode. This
  situation is reminiscent of the ``breathing'' memory in
  scalar-tensor theories of gravity
  \cite{Campiglia:2018see,Seraj:2021qja}. We leave this problem to
  future work.}
		
These equations can be inverted to compute the memory fields using the
following Green functions
\begin{subequations}
\label{Green functions}  
\begin{align}
    G_1(\nb,\nb')&=\tfrac{1}{4\pi}\ln x&    G_2(\nb,\nb')&=\tfrac{1}{4\pi}x\ln x\,,
\end{align}
\end{subequations}
with $x\equiv 1-\nb\cdot \nb'$, which respectively solve
\begin{align*}
    D^2 G_1(\nb,\nb')&=\tfrac{1}{\sqrt{\gamma}}\delta^2(\th,\th')-\tfrac{1}{4\pi}\\
    \tfrac12 D^2(D^2+2) G_2(\nb,\nb')&=\tfrac{1}{\sqrt{\gamma}}\delta^2(\th,\th')-\tfrac{1}{4\pi}(1+3\nb\cdot\nb') \, .
\end{align*}
As a result, vacuum fields are given by 
\begin{subequations}
\begin{align}
    C_\mathrm{V}(\nb') & =\frac{{n'}{}^i}{4\pi}\int d\Omega\,  \frac{\delta\bA{}^{\mathrm{TL}}_i(\nb)}{1-\nb\cdot \nb'}  \\
    C_\mathrm{T}(\nb') & =\frac{{n'}{}^i{n'}{}^j}{4\pi}\int d\Omega   \, \frac{\delta{\bA}{}^{\mathrm{TT}}_{ij}(\nb)}{1-\nb\cdot \nb'} \\
    C_\mathrm{S}(\nb) & =\bP{}^{ij}\delta \bA{}^{\mathrm{T}}_{ij}(\nb)\,,
\end{align}
\end{subequations}
where $A\equiv \bP{}^{ij}{\bA}{}^{T}_{ij}$ and
${\bA}{}^\mathrm{TT}_{ij}={\bA}{}^\mathrm{T}_{ij}-\frac{1}{2}\bP_{ij}A$.
Implementing \eqref{eq:memory_dip} in these equations reveals the
memory fields. The change in the charges resulting from radiation flux
is also exclusively determined by the memory fields.
Integrating~\eqref{eq:fluxTT} and~\eqref{eq:FluxTL} in time, and
using~\eqref{eq:DeltaA_Gold}, one finds that the flux of charges
through the fast radiation ($\mathrm{T}$-sector) is given by
\begin{subequations}
  \begin{align}
    \delta\mathcal{P} & =-\tfrac{1}{2}c\sqrt{\gamma}\left(D^{2}+2\right)\left(D^{2} C_\mathrm{T}+2C_\mathrm{S}\right) \\
    \delta\mathcal{Q}& =2c\sqrt{\gamma}\,C_\mathrm{S} \, ,
  \end{align}
\end{subequations}
while the slow radiation ($\mathrm{TL}$-sector) carries
\begin{align}
  \delta\mathcal{Q} & =-2\ct\sqrt{\gamma}\,D^{2}C_\mathrm{V}& \delta\mathcal{P} & =0 \, .
\end{align}
Thus we have established connections between asymptotic symmetries and
memory effects in the fracton infrared triangle (see
Figure~\ref{fig:frac-triangle}).
            
\section{Soft factors from the memory effect}
\label{sec:soft-factor-from}

In this section we determine the soft factors for a scattering process
of dipoles from the memory effect.

Consider a scattering process of $N$ dipoles with momentum
$\pb_{\alpha}=m_{\alpha} \vb_{\alpha}$ and dipole moment
$\db_{\alpha}$, with the emission of one fractonic soft photon with
frequency $\omega$ and polarization projectors $\epsilon^{ij}_{\#}$.
The scattering amplitude is expected to factorize in the soft limit
$\omega\to 0$ as
\begin{align}\label{soft theorem}
\mathcal{A}_{n+1}\big(\vb_{\alpha},\db_{\alpha};\omega,\epsilon^{ij}_{\#}\big)=\frac{1}{\omega}\epsilon^{ij}_{\#}S_{ij}^{\#}\;\mathcal{A}_{n}\big(\vb_{\alpha},\db_{\alpha}\big)+O(\omega^0) \, .
\end{align}

To derive the soft factor $S_{ij}^{\#}$, we will closely
follow~\cite{Strominger:2014pwa}. We illustrate the derivation for the
T sector, while the analysis for the TL sector would be similar.
Starting from a Fourier mode decomposition of the gauge field
$A^{\rm T}_{ij}$, it can be shown that the radiative field
$\bA{}^{\rm T}_{ij}=\lim_{r\to\infty}(r\bA{}^{\rm T}_{ij})$ can be
computed using a saddle point approximation
\begin{align}
    \bA{}^{\rm T}_{ij}(u,\bm{n})=\frac{i}{2}\int \frac{d\omega}{(2\pi c)^2} \Big(\epsilon_{ij}^\alpha \,a_\alpha^\dagger(\omega, {\bm n}) e^{i\omega u}-{\rm c.c.}\Big)
\end{align}
where $\alpha$ labels the transverse mode with polarization
$\epsilon_{ij}^\alpha$, created by $a_\alpha^\dagger$. The next step
is to use this result to compute the memory
\begin{align}
  \delta \bA{}^{\rm T}_{ij}=\int du {\dot\bA}{}^{\rm T}_{ij}=-\frac{1}{4\pi c^2}\epsilon^\alpha_{ij}\lim_{\omega\to 0} \big(\omega a^\dagger_\alpha(\omega,{\bm n})+ {\rm c.c.} \big) \, .
\end{align}

This equation relates the memory to the creation and annihilation of a
fractonic soft (zero frequency) photon. As a result, an amplitude with
an external soft photon factorizes according to \eqref{soft theorem}
with soft factors
\begin{equation}
  \label{eq:soft_factors}
    S_{ij}^{\#}=-4\pi c_{\#}^{2}\delta \overline{A}_{ij}^{\#} \, .
\end{equation}
Thus the soft factors $S_{ij}^{\mathrm{T}},S_{ij}^{\mathrm{TL}}$ in
\eqref{soft theorem} are given up to overall factors
$-4\pi c_{\#}^{2}$ by ~\eqref{memory T} and \eqref{memory TL}
respectively. An alternative derivation of the soft factors is
detailed in the supplemental material and involves the use of Feynman
diagrams within a simple effective model that describes the dynamics
of dipoles coupled to the fracton gauge field. The derivation of the
soft factor from the conservation of asymptotic charges will be
discussed in our extended work~\cite{Perez:2024xxx}.
		
\section{Discussion and outlook}
\label{sec:discussion}
		
In this work, we introduced an observable double kick memory effect
(Section~\ref{sec:Radiation}) and the corresponding dipole soft
theorem (Section~\ref{sec:soft-factor-from}), which we related to the
asymptotic symmetries of fracton gauge theory. This provides the first
instance of an IR triangle (Figure~\ref{fig:frac-triangle}) for a
theory beyond Lorentz symmetry and further evidence for the
robustness of this triangular correspondence.
		
The tools developed and implemented in this work can be used to study
radiation and IR effects beyond lorentzian symmetries, which opens the
door to explore other models of relevance to condensed matter systems
and beyond, e.g.,~\cite{Xu_2006,2016arXiv160108235R,Bulmash:2018knk,
  Pretko:2016kxt,Pretko:2016lgv,Bulmash:2018knk,
  Schmitz:2018kbo,Slagle:2018swq,Gromov:2018nbv, Li:2019tje,
  Prem:2019etl,Radzihovsky:2019jdo,Seiberg:2019vrp,Wang:2019aiq,
  Shenoy:2019wng,Radicevic:2019vyb,Wang:2019cbj,
  Argurio:2021opr,Casalbuoni:2021fel,
  Pena-Benitez:2021ipo,Angus:2021jvm, Lake:2022ico, Jensen:2022iww,
  Brauner:2023mne, Baig:2023yaz,Kasikci:2023tvs,
  Cheung:2023qwn,Molina-Vilaplana:2023doq,
  Bertolini:2023sqa,Ebisu:2023idd, Pena-Benitez:2023aat}. Conversely,
studying theories beyond Lorentz symmetry can improve our
understanding of IR structures (similar to Nambu--Goldstone modes
whose dispersion relations are richer and more intricate once Lorentz
symmetry is not imposed). The double kick effect (see
Figure~\ref{fig:double-kick}) is an example of a novel memory
observable which exhibits the more intricate structure that can appear
in a non-lorentzian theory.

One avenue we would like to highlight involves leveraging the duality
between (specific models of) fractons and
vortices~\cite{Doshi:2020jso} as well as the relationship with
elasticity~\cite{Pretko:2017kvd, Pretko:2019omh}
(see~\cite{Gromov:2022cxa} for a review). These dualities open up
exciting possibilities for creating experimental setups that are
potentially easier to realize and which could facilitate the
observation of memory effects (see also \cite{Tsaloukidis:2023jmr}).

While fractons have originated from condensed matter physics, they
might play an important role in the holographic understanding of
gravity in asymptotically flat spacetimes. The reason is the
correspondence between the fracton algebra and the Carroll algebra
which is the underlying symmetry of gravity in asymptotically flat
spacetimes~\cite{Figueroa-OFarrill:2023vbj,Figueroa-OFarrill:2023qty,Duval:2014uoa,Bergshoeff:2014jla,Marsot:2022imf,Zhang:2023jbi}.
Indeed, many of the discussed dualities and experiments can equally
well be seen though the lens of carrollian physics, e.g.,
in~\cite{Armas:2023dcz} insights from fractons have been used in the
context of Carroll fluids.


In this work, we have focused on the leading IR behavior. However, it
might be interesting to explore subleading effects and consider how
celestial
holography~\cite{Raclariu:2021zjz,Pasterski:2021rjz,Pasterski:2021raf,McLoughlin:2022ljp,Donnay:2023mrd}
could be extended to this setup. After all Lorentz symmetry is absent,
but we still recover an analog IR structure. In addition, to get
closer to experiments that investigate dipole symmetry
\cite{Guardado-Sanchez:2019bjm,Scherg:2020mcp,2021arXiv210615586K,2022PhRvX..12b1014Z}
it might also be interesting to introduce and study boundaries at
finite distances.

Motivated by the historically prolific interdisciplinary dialogue
between high-energy and condensed matter physics, we are also excited
by the prospect that the inaugural experimental validation of memory
effects may manifest within the domain of condensed matter systems. We
hope that this work will serve as an initial stepping stone for this
promising endeavor.
\begin{acknowledgments}
	The research of AP is partially supported by Fondecyt grants No
	1211226, 1220910 and 1230853. SP was supported by the Leverhulme
	Trust Research Project Grant (RPG-2019-218) ``What is
	Non-Relativistic Quantum Gravity and is it Holographic?''.
\end{acknowledgments}

\providecommand{\href}[2]{#2}\begingroup\raggedright\endgroup

\begin{thebibliography}{10}

\bibitem{Chamon:2004lew}
C.~Chamon, ``{Quantum Glassiness},''
  \href{http://dx.doi.org/10.1103/physrevlett.94.040402}{{\em Phys. Rev. Lett.}
  {\bfseries 94} no.~4, (2005) 040402},
  \href{http://arxiv.org/abs/cond-mat/0404182}{{\ttfamily
  arXiv:cond-mat/0404182}}.

\bibitem{Haah:2011drr}
J.~Haah, ``{Local stabilizer codes in three dimensions without string logical
  operators},'' \href{http://dx.doi.org/10.1103/physreva.83.042330}{{\em Phys.
  Rev. A} {\bfseries 83} no.~4, (2011) 042330},
  \href{http://arxiv.org/abs/1101.1962}{{\ttfamily arXiv:1101.1962
  [quant-ph]}}.

\bibitem{Nandkishore:2018sel}
R.~M. Nandkishore and M.~Hermele, ``{Fractons},''
  \href{http://dx.doi.org/10.1146/annurev-conmatphys-031218-013604}{{\em Ann.
  Rev. Condensed Matter Phys.} {\bfseries 10} (2019) 295--313},
  \href{http://arxiv.org/abs/1803.11196}{{\ttfamily arXiv:1803.11196
  [cond-mat.str-el]}}.

\bibitem{Pretko:2020cko}
M.~Pretko, X.~Chen, and Y.~You, ``{Fracton Phases of Matter},''
  \href{http://dx.doi.org/10.1142/S0217751X20300033}{{\em Int. J. Mod. Phys. A}
  {\bfseries 35} no.~06, (2020) 2030003},
  \href{http://arxiv.org/abs/2001.01722}{{\ttfamily arXiv:2001.01722
  [cond-mat.str-el]}}.

\bibitem{Grosvenor:2021hkn}
K.~T. Grosvenor, C.~Hoyos, F.~Pe\~na Benitez, and P.~Sur\'owka,
  ``{Space-Dependent Symmetries and Fractons},''
  \href{http://dx.doi.org/10.3389/fphy.2021.792621}{{\em Front. in Phys.}
  {\bfseries 9} (2022) 792621},
  \href{http://arxiv.org/abs/2112.00531}{{\ttfamily arXiv:2112.00531
  [hep-th]}}.

\bibitem{Brauner:2022rvf}
T.~Brauner, S.~A. Hartnoll, P.~Kovtun, H.~Liu, M.~Mezei, A.~Nicolis, R.~Penco,
  S.-H. Shao, and D.~T. Son, ``{Snowmass White Paper: Effective Field Theories
  for Condensed Matter Systems},'' in {\em {2022 Snowmass Summer Study}}.
\newblock 3, 2022.
\newblock \href{http://arxiv.org/abs/2203.10110}{{\ttfamily arXiv:2203.10110
  [hep-th]}}.

\bibitem{Cordova:2022ruw}
C.~Cordova, T.~T. Dumitrescu, K.~Intriligator, and S.-H. Shao, ``{Snowmass
  White Paper: Generalized Symmetries in Quantum Field Theory and Beyond},'' in
  {\em {2022 Snowmass Summer Study}}.
\newblock 5, 2022.
\newblock \href{http://arxiv.org/abs/2205.09545}{{\ttfamily arXiv:2205.09545
  [hep-th]}}.

\bibitem{Bidussi:2021nmp}
L.~Bidussi, J.~Hartong, E.~Have, J.~Musaeus, and S.~Prohazka, ``{Fractons,
  dipole symmetries and curved spacetime},''
  \href{http://dx.doi.org/10.21468/SciPostPhys.12.6.205}{{\em SciPost Phys.}
  {\bfseries 12} no.~6, (2022) 205},
  \href{http://arxiv.org/abs/2111.03668}{{\ttfamily arXiv:2111.03668
  [hep-th]}}.

\bibitem{Marsot:2022imf}
L.~Marsot, P.~M. Zhang, M.~Chernodub, and P.~A. Horvathy, ``{Hall effects in
  Carroll dynamics},''
  \href{http://dx.doi.org/10.1016/j.physrep.2023.07.007}{{\em Phys. Rept.}
  {\bfseries 1028} (2023) 1--60},
  \href{http://arxiv.org/abs/2212.02360}{{\ttfamily arXiv:2212.02360
  [hep-th]}}.

\bibitem{Figueroa-OFarrill:2023vbj}
J.~Figueroa-O'Farrill, A.~P\'erez, and S.~Prohazka, ``{Carroll/fracton
  particles and their correspondence},''
  \href{http://dx.doi.org/10.1007/JHEP06(2023)207}{{\em JHEP} {\bfseries 06}
  (5, 2023) 207}, \href{http://arxiv.org/abs/2305.06730}{{\ttfamily
  arXiv:2305.06730 [hep-th]}}.

\bibitem{Figueroa-OFarrill:2023qty}
J.~Figueroa-O'Farrill, A.~P\'erez, and S.~Prohazka, ``{Quantum Carroll/fracton
  particles},'' \href{http://dx.doi.org/10.1007/JHEP10(2023)041}{{\em JHEP}
  {\bfseries 10} (2023) 041}, \href{http://arxiv.org/abs/2307.05674}{{\ttfamily
  arXiv:2307.05674 [hep-th]}}.

\bibitem{Figueroa-OFarrill:2021sxz}
J.~Figueroa-O'Farrill, E.~Have, S.~Prohazka, and J.~Salzer, ``{Carrollian and
  celestial spaces at infinity},''
  \href{http://dx.doi.org/10.1007/JHEP09(2022)007}{{\em JHEP} {\bfseries 09}
  (2022) 007}, \href{http://arxiv.org/abs/2112.03319}{{\ttfamily
  arXiv:2112.03319 [hep-th]}}.

\bibitem{Donnay:2022aba}
L.~Donnay, A.~Fiorucci, Y.~Herfray, and R.~Ruzziconi, ``{Carrollian Perspective
  on Celestial Holography},''
  \href{http://dx.doi.org/10.1103/PhysRevLett.129.071602}{{\em Phys. Rev.
  Lett.} {\bfseries 129} no.~7, (2022) 071602},
  \href{http://arxiv.org/abs/2202.04702}{{\ttfamily arXiv:2202.04702
  [hep-th]}}.

\bibitem{Bagchi:2022emh}
A.~Bagchi, S.~Banerjee, R.~Basu, and S.~Dutta, ``{Scattering Amplitudes:
  Celestial and Carrollian},''
  \href{http://dx.doi.org/10.1103/PhysRevLett.128.241601}{{\em Phys. Rev.
  Lett.} {\bfseries 128} no.~24, (2022) 241601},
  \href{http://arxiv.org/abs/2202.08438}{{\ttfamily arXiv:2202.08438
  [hep-th]}}.

\bibitem{Donnay:2022wvx}
L.~Donnay, A.~Fiorucci, Y.~Herfray, and R.~Ruzziconi, ``{Bridging Carrollian
  and Celestial Holography},''
  \href{http://arxiv.org/abs/2212.12553}{{\ttfamily arXiv:2212.12553
  [hep-th]}}.

\bibitem{Strominger:2017zoo}
A.~Strominger, {\em {Lectures on the Infrared Structure of Gravity and Gauge
  Theory}}.
\newblock Princeton University Press, 2018.
\newblock
\href{http://arxiv.org/abs/1703.05448}{{\ttfamily arXiv:1703.05448 [hep-th]}}.
\newblock

\bibitem{Raclariu:2021zjz}
A.-M. Raclariu, ``{Lectures on Celestial Holography},''
  \href{http://arxiv.org/abs/2107.02075}{{\ttfamily arXiv:2107.02075
  [hep-th]}}.

\bibitem{Pasterski:2021rjz}
S.~Pasterski, ``{Lectures on celestial amplitudes},''
  \href{http://dx.doi.org/10.1140/epjc/s10052-021-09846-7}{{\em Eur. Phys. J.
  C} {\bfseries 81} no.~12, (2021) 1062},
  \href{http://arxiv.org/abs/2108.04801}{{\ttfamily arXiv:2108.04801
  [hep-th]}}.

\bibitem{Pasterski:2021raf}
S.~Pasterski, M.~Pate, and A.-M. Raclariu, ``{Celestial Holography},'' in {\em
  {2022 Snowmass Summer Study}}.
\newblock 11, 2021.
\newblock \href{http://arxiv.org/abs/2111.11392}{{\ttfamily arXiv:2111.11392
  [hep-th]}}.

\bibitem{McLoughlin:2022ljp}
T.~McLoughlin, A.~Puhm, and A.-M. Raclariu, ``{The SAGEX Review on Scattering
  Amplitudes, Chapter 11: Soft Theorems and Celestial Amplitudes},''
  \href{http://arxiv.org/abs/2203.13022}{{\ttfamily arXiv:2203.13022
  [hep-th]}}.

\bibitem{Donnay:2023mrd}
L.~Donnay, ``{Celestial holography: An asymptotic symmetry perspective},''
  \href{http://arxiv.org/abs/2310.12922}{{\ttfamily arXiv:2310.12922
  [hep-th]}}.

\bibitem{Zeldovich:1974gvh}
Y.~B. Zel'dovich and A.~G. Polnarev, ``{Radiation of gravitational waves by a
  cluster of superdense stars},'' {\em Sov. Astron.} {\bfseries 18} (1974) 17.

\bibitem{Braginsky:1985vlg}
V.~B. Braginsky and L.~P. Grishchuk, ``{Kinematic Resonance and Memory Effect
  in Free Mass Gravitational Antennas},'' {\em Sov. Phys. JETP} {\bfseries 62}
  (1985) 427--430.

\bibitem{braginsky1987gravitational}
V.~B. Braginsky and K.~S. Thorne, ``Gravitational-wave bursts with memory and
  experimental prospects,'' {\em Nature} {\bfseries 327} no.~6118, (1987)
  123--125.

\bibitem{Christodoulou:1991cr}
D.~Christodoulou, ``{Nonlinear nature of gravitation and gravitational wave
  experiments},'' \href{http://dx.doi.org/10.1103/PhysRevLett.67.1486}{{\em
  Phys. Rev. Lett.} {\bfseries 67} (1991) 1486--1489}.

\bibitem{Strominger:2013lka}
A.~Strominger, ``{Asymptotic Symmetries of Yang-Mills Theory},''
  \href{http://dx.doi.org/10.1007/JHEP07(2014)151}{{\em JHEP} {\bfseries 07}
  (2014) 151},
\href{http://arxiv.org/abs/1308.0589}{{\ttfamily arXiv:1308.0589 [hep-th]}}.

\bibitem{Strominger:2014pwa}
A.~Strominger and A.~Zhiboedov, ``{Gravitational Memory, BMS Supertranslations
  and Soft Theorems},'' \href{http://dx.doi.org/10.1007/JHEP01(2016)086}{{\em
  JHEP} {\bfseries 01} (2016) 086},
\href{http://arxiv.org/abs/1411.5745}{{\ttfamily arXiv:1411.5745 [hep-th]}}.

\bibitem{Strominger:2015bla}
A.~Strominger, ``{Magnetic Corrections to the Soft Photon Theorem},''
  \href{http://dx.doi.org/10.1103/PhysRevLett.116.031602}{{\em Phys. Rev.
  Lett.} {\bfseries 116} no.~3, (2016) 031602},
\href{http://arxiv.org/abs/1509.00543}{{\ttfamily arXiv:1509.00543 [hep-th]}}.

\bibitem{Bieri:2013ada}
L.~Bieri and D.~Garfinkle, ``{Perturbative and gauge invariant treatment of
  gravitational wave memory},''
  \href{http://dx.doi.org/10.1103/PhysRevD.89.084039}{{\em Phys. Rev. D}
  {\bfseries 89} no.~8, (2014) 084039},
  \href{http://arxiv.org/abs/1312.6871}{{\ttfamily arXiv:1312.6871 [gr-qc]}}.

\bibitem{Bieri:2013hqa}
L.~Bieri and D.~Garfinkle, ``{An electromagnetic analogue of gravitational wave
  memory},'' \href{http://dx.doi.org/10.1088/0264-9381/30/19/195009}{{\em
  Class. Quant. Grav.} {\bfseries 30} (2013) 195009},
  \href{http://arxiv.org/abs/1307.5098}{{\ttfamily arXiv:1307.5098 [gr-qc]}}.

\bibitem{Pasterski:2015tva}
S.~Pasterski, A.~Strominger, and A.~Zhiboedov, ``{New Gravitational
  Memories},'' \href{http://dx.doi.org/10.1007/JHEP12(2016)053}{{\em JHEP}
  {\bfseries 12} (2016) 053}, \href{http://arxiv.org/abs/1502.06120}{{\ttfamily
  arXiv:1502.06120 [hep-th]}}.

\bibitem{Pasterski:2015zua}
S.~Pasterski, ``{Asymptotic Symmetries and Electromagnetic Memory},''
  \href{http://dx.doi.org/10.1007/JHEP09(2017)154}{{\em JHEP} {\bfseries 09}
  (2017) 154}, \href{http://arxiv.org/abs/1505.00716}{{\ttfamily
  arXiv:1505.00716 [hep-th]}}.

\bibitem{Flanagan:2019ezo}
E.~E. Flanagan, A.~M. Grant, A.~I. Harte, and D.~A. Nichols, ``{Persistent
  gravitational wave observables: Nonlinear plane wave spacetimes},''
  \href{http://dx.doi.org/10.1103/PhysRevD.101.104033}{{\em Phys. Rev. D}
  {\bfseries 101} no.~10, (2020) 104033},
  \href{http://arxiv.org/abs/1912.13449}{{\ttfamily arXiv:1912.13449 [gr-qc]}}.

\bibitem{Nichols:2018qac}
D.~A. Nichols, ``{Center-of-mass angular momentum and memory effect in
  asymptotically flat spacetimes},''
  \href{http://dx.doi.org/10.1103/PhysRevD.98.064032}{{\em Phys. Rev. D}
  {\bfseries 98} no.~6, (2018) 064032},
  \href{http://arxiv.org/abs/1807.08767}{{\ttfamily arXiv:1807.08767 [gr-qc]}}.

\bibitem{Seraj:2021qja}
A.~Seraj, ``{Gravitational breathing memory and dual symmetries},''
  \href{http://dx.doi.org/10.1007/JHEP05(2021)283}{{\em JHEP} {\bfseries 05}
  (2021) 283}, \href{http://arxiv.org/abs/2103.12185}{{\ttfamily
  arXiv:2103.12185 [hep-th]}}.

\bibitem{Seraj:2021rxd}
A.~Seraj and B.~Oblak, ``{Gyroscopic gravitational memory},''
  \href{http://dx.doi.org/10.1007/JHEP11(2023)057}{{\em JHEP} {\bfseries 11}
  (2023) 057}, \href{http://arxiv.org/abs/2112.04535}{{\ttfamily
  arXiv:2112.04535 [hep-th]}}.

\bibitem{Seraj:2022qyt}
A.~Seraj and B.~Oblak, ``{Precession Caused by Gravitational Waves},''
  \href{http://dx.doi.org/10.1103/PhysRevLett.129.061101}{{\em Phys. Rev.
  Lett.} {\bfseries 129} no.~6, (2022) 061101},
  \href{http://arxiv.org/abs/2203.16216}{{\ttfamily arXiv:2203.16216 [gr-qc]}}.

\bibitem{Seraj:2022qqj}
A.~Seraj and T.~Neogi, ``{Memory effects from holonomies},''
  \href{http://arxiv.org/abs/2206.14110}{{\ttfamily arXiv:2206.14110
  [hep-th]}}.

\bibitem{Oblak:2023axy}
B.~Oblak and A.~Seraj, ``{Orientation Memory of Magnetic Dipoles},''
  \href{http://arxiv.org/abs/2304.12348}{{\ttfamily arXiv:2304.12348
  [hep-th]}}.

\bibitem{Guardado-Sanchez:2019bjm}
E.~Guardado-Sanchez, A.~Morningstar, B.~M. Spar, P.~T. Brown, D.~A. Huse, and
  W.~S. Bakr, ``{Subdiffusion and Heat Transport in a Tilted Two-Dimensional
  Fermi-Hubbard System},''
  \href{http://dx.doi.org/10.1103/PhysRevX.10.011042}{{\em Phys. Rev. X}
  {\bfseries 10} no.~1, (2020) 011042},
  \href{http://arxiv.org/abs/1909.05848}{{\ttfamily arXiv:1909.05848
  [cond-mat.quant-gas]}}.

\bibitem{Scherg:2020mcp}
S.~Scherg, T.~Kohlert, P.~Sala, F.~Pollmann, B.~Hebbe~Madhusudhana, I.~Bloch,
  and M.~Aidelsburger, ``{Observing non-ergodicity due to kinetic constraints
  in tilted Fermi-Hubbard chains},''
  \href{http://dx.doi.org/10.1038/s41467-021-24726-0}{{\em Nature Commun.}
  {\bfseries 12} no.~1, (2021) 4490},
  \href{http://arxiv.org/abs/2010.12965}{{\ttfamily arXiv:2010.12965
  [cond-mat.quant-gas]}}.

\bibitem{2021arXiv210615586K}
T.~{Kohlert}, S.~{Scherg}, P.~{Sala}, F.~{Pollmann}, B.~{Hebbe Madhusudhana},
  I.~{Bloch}, and M.~{Aidelsburger}, ``{Experimental realization of fragmented
  models in tilted Fermi-Hubbard chains},''
  \href{http://dx.doi.org/10.48550/arXiv.2106.15586}{{\em arXiv e-prints}
  (June, 2021) arXiv:2106.15586},
  \href{http://arxiv.org/abs/2106.15586}{{\ttfamily arXiv:2106.15586
  [cond-mat.quant-gas]}}.

\bibitem{2022PhRvX..12b1014Z}
H.~P. {Zahn}, V.~P. {Singh}, M.~N. {Kosch}, L.~{Asteria}, L.~{Freystatzky},
  K.~{Sengstock}, L.~{Mathey}, and C.~{Weitenberg}, ``{Formation of Spontaneous
  Density-Wave Patterns in dc Driven Lattices},''
  \href{http://dx.doi.org/10.1103/PhysRevX.12.021014}{{\em Physical Review X}
  {\bfseries 12} no.~2, (Apr., 2022) 021014},
  \href{http://arxiv.org/abs/2108.11917}{{\ttfamily arXiv:2108.11917
  [cond-mat.quant-gas]}}.

\bibitem{Perez:2024xxx}
A.~P\'erez, S.~Prohazka, and A.~Seraj, ``{Fracton radiation, asymptotic
  structure and the infrared triangle beyond lorentz symmetry (in
  preparation)},'' \href{http://arxiv.org/abs/24XX.XXXXX}{{\ttfamily
  arXiv:24XX.XXXXX [hep-th]}}.

\bibitem{Pretko:2016kxt}
M.~Pretko, ``{Subdimensional Particle Structure of Higher Rank U(1) Spin
  Liquids},'' \href{http://dx.doi.org/10.1103/PhysRevB.95.115139}{{\em Phys.
  Rev. B} {\bfseries 95} no.~11, (2017) 115139},
  \href{http://arxiv.org/abs/1604.05329}{{\ttfamily arXiv:1604.05329
  [cond-mat.str-el]}}.

\bibitem{Pretko:2016lgv}
M.~Pretko, ``{Generalized Electromagnetism of Subdimensional Particles: A Spin
  Liquid Story},'' \href{http://dx.doi.org/10.1103/PhysRevB.96.035119}{{\em
  Phys. Rev. B} {\bfseries 96} no.~3, (2017) 035119},
  \href{http://arxiv.org/abs/1606.08857}{{\ttfamily arXiv:1606.08857
  [cond-mat.str-el]}}.

\bibitem{Jain:2021ibh}
A.~Jain and K.~Jensen, ``{Fractons in curved space},''
  \href{http://dx.doi.org/10.21468/SciPostPhys.12.4.142}{{\em SciPost Phys.}
  {\bfseries 12} no.~4, (2022) 142},
  \href{http://arxiv.org/abs/2111.03973}{{\ttfamily arXiv:2111.03973
  [hep-th]}}.

\bibitem{Perez:2022kax}
A.~P\'erez and S.~Prohazka, ``{Asymptotic symmetries and soft charges of
  fractons},'' \href{http://dx.doi.org/10.1103/PhysRevD.106.044017}{{\em Phys.
  Rev. D} {\bfseries 106} no.~4, (2022) 044017},
  \href{http://arxiv.org/abs/2203.02817}{{\ttfamily arXiv:2203.02817
  [hep-th]}}.

\bibitem{Regge:1974zd}
T.~Regge and C.~Teitelboim, ``{Role of Surface Integrals in the Hamiltonian
  Formulation of General Relativity},''
\href{http://dx.doi.org/10.1016/0003-4916(74)90404-7}{{\em Annals Phys.}
  {\bfseries 88} (1974) 286}.

\bibitem{Lee:1990nz}
J.~Lee and R.~M. Wald, ``{Local symmetries and constraints},''
  \href{http://dx.doi.org/10.1063/1.528801}{{\em J. Math. Phys.} {\bfseries 31}
  (1990) 725--743}.

\bibitem{Bondi:1960jsa}
H.~Bondi, ``{Gravitational Waves in General Relativity},''
  \href{http://dx.doi.org/10.1038/186535a0}{{\em Nature} {\bfseries 186}
  no.~4724, (1960) 535--535}.

\bibitem{Bondi:1962px}
H.~Bondi, M.~G.~J. van~der Burg, and A.~W.~K. Metzner, ``{Gravitational waves
  in general relativity. 7. Waves from axisymmetric isolated systems},''
\href{http://dx.doi.org/10.1098/rspa.1962.0161}{{\em Proc. Roy. Soc. Lond.}
  {\bfseries A269} (1962) 21--52}.

\bibitem{Campiglia:2018see}
M.~Campiglia, L.~Freidel, F.~Hopfmueller, and R.~M. Soni, ``{Scalar Asymptotic
  Charges and Dual Large Gauge Transformations},''
  \href{http://dx.doi.org/10.1007/JHEP04(2019)003}{{\em JHEP} {\bfseries 04}
  (2019) 003}, \href{http://arxiv.org/abs/1810.04213}{{\ttfamily
  arXiv:1810.04213 [hep-th]}}.

\bibitem{Xu_2006}
C.~Xu, ``Gapless bosonic excitation without symmetry breaking: An algebraic
  spin liquid with soft gravitons,''
  \href{http://dx.doi.org/10.1103/physrevb.74.224433}{{\em Physical Review B}
  {\bfseries 74} no.~22, (Dec, 2006) }.
  \url{http://dx.doi.org/10.1103/PhysRevB.74.224433}.

\bibitem{2016arXiv160108235R}
A.~{Rasmussen}, Y.-Z. {You}, and C.~{Xu}, ``{Stable Gapless Bose Liquid Phases
  without any Symmetry},''
  \href{http://dx.doi.org/10.48550/arXiv.1601.08235}{{\em arXiv e-prints}
  (Jan., 2016) arXiv:1601.08235},
  \href{http://arxiv.org/abs/1601.08235}{{\ttfamily arXiv:1601.08235
  [cond-mat.str-el]}}.

\bibitem{Bulmash:2018knk}
D.~Bulmash and M.~Barkeshli, ``{Generalized $U(1)$ Gauge Field Theories and
  Fractal Dynamics},'' \href{http://arxiv.org/abs/1806.01855}{{\ttfamily
  arXiv:1806.01855 [cond-mat.str-el]}}.

\bibitem{Schmitz:2018kbo}
A.~T. Schmitz, ``{Gauge Structures: From Stabilizer Codes to Continuum
  Models},'' \href{http://dx.doi.org/10.1016/j.aop.2019.167927}{{\em Annals
  Phys.} {\bfseries 410} (2019) 167927},
  \href{http://arxiv.org/abs/1809.10151}{{\ttfamily arXiv:1809.10151
  [quant-ph]}}.

\bibitem{Slagle:2018swq}
K.~Slagle, D.~Aasen, and D.~Williamson, ``{Foliated Field Theory and
  String-Membrane-Net Condensation Picture of Fracton Order},''
  \href{http://dx.doi.org/10.21468/SciPostPhys.6.4.043}{{\em SciPost Phys.}
  {\bfseries 6} no.~4, (2019) 043},
  \href{http://arxiv.org/abs/1812.01613}{{\ttfamily arXiv:1812.01613
  [cond-mat.str-el]}}.

\bibitem{Gromov:2018nbv}
A.~Gromov, ``{Towards classification of Fracton phases: the multipole
  algebra},'' \href{http://dx.doi.org/10.1103/PhysRevX.9.031035}{{\em Phys.
  Rev. X} {\bfseries 9} no.~3, (2019) 031035},
  \href{http://arxiv.org/abs/1812.05104}{{\ttfamily arXiv:1812.05104
  [cond-mat.str-el]}}.

\bibitem{Li:2019tje}
M.-Y. Li and P.~Ye, ``{Fracton physics of spatially extended excitations},''
  \href{http://dx.doi.org/10.1103/PhysRevB.101.245134}{{\em Phys. Rev. B}
  {\bfseries 101} no.~24, (2020) 245134},
  \href{http://arxiv.org/abs/1909.02814}{{\ttfamily arXiv:1909.02814
  [cond-mat.str-el]}}.

\bibitem{Prem:2019etl}
A.~Prem and D.~J. Williamson, ``{Gauging permutation symmetries as a route to
  non-Abelian fractons},''
  \href{http://dx.doi.org/10.21468/SciPostPhys.7.5.068}{{\em SciPost Phys.}
  {\bfseries 7} no.~5, (2019) 068},
  \href{http://arxiv.org/abs/1905.06309}{{\ttfamily arXiv:1905.06309
  [cond-mat.str-el]}}.

\bibitem{Radzihovsky:2019jdo}
L.~Radzihovsky and M.~Hermele, ``{Fractons from vector gauge theory},''
  \href{http://dx.doi.org/10.1103/PhysRevLett.124.050402}{{\em Phys. Rev.
  Lett.} {\bfseries 124} no.~5, (2020) 050402},
  \href{http://arxiv.org/abs/1905.06951}{{\ttfamily arXiv:1905.06951
  [cond-mat.str-el]}}.

\bibitem{Seiberg:2019vrp}
N.~Seiberg, ``{Field Theories With a Vector Global Symmetry},''
  \href{http://dx.doi.org/10.21468/SciPostPhys.8.4.050}{{\em SciPost Phys.}
  {\bfseries 8} no.~4, (2020) 050},
  \href{http://arxiv.org/abs/1909.10544}{{\ttfamily arXiv:1909.10544
  [cond-mat.str-el]}}.

\bibitem{Wang:2019aiq}
J.~Wang and K.~Xu, ``{Higher-Rank Tensor Field Theory of Non-Abelian Fracton
  and Embeddon},'' \href{http://dx.doi.org/10.1016/j.aop.2020.168370}{{\em
  Annals Phys.} {\bfseries 424} (2021) 168370},
  \href{http://arxiv.org/abs/1909.13879}{{\ttfamily arXiv:1909.13879
  [hep-th]}}.

\bibitem{Shenoy:2019wng}
V.~B. Shenoy and R.~Moessner, ``{$(k,n)$-fractonic Maxwell theory},''
  \href{http://dx.doi.org/10.1103/PhysRevB.101.085106}{{\em Phys. Rev. B}
  {\bfseries 101} no.~8, (2020) 085106},
  \href{http://arxiv.org/abs/1910.02820}{{\ttfamily arXiv:1910.02820
  [cond-mat.str-el]}}.

\bibitem{Radicevic:2019vyb}
D.~Radi\v{c}evi\'c, ``{Systematic Constructions of Fracton Theories},''
  \href{http://arxiv.org/abs/1910.06336}{{\ttfamily arXiv:1910.06336
  [cond-mat.str-el]}}.

\bibitem{Wang:2019cbj}
J.~Wang, K.~Xu, and S.-T. Yau, ``{Higher-rank tensor non-Abelian field theory:
  Higher-moment or subdimensional polynomial global symmetry, algebraic
  variety, Noether's theorem, and gauging},''
  \href{http://dx.doi.org/10.1103/PhysRevResearch.3.013185}{{\em Phys. Rev.
  Res.} {\bfseries 3} no.~1, (2021) 013185},
  \href{http://arxiv.org/abs/1911.01804}{{\ttfamily arXiv:1911.01804
  [hep-th]}}.

\bibitem{Argurio:2021opr}
R.~Argurio, C.~Hoyos, D.~Musso, and D.~Naegels, ``{Fractons in effective field
  theories for spontaneously broken translations},''
  \href{http://dx.doi.org/10.1103/PhysRevD.104.105001}{{\em Phys. Rev. D}
  {\bfseries 104} no.~10, (2021) 105001},
  \href{http://arxiv.org/abs/2107.03073}{{\ttfamily arXiv:2107.03073
  [hep-th]}}.

\bibitem{Casalbuoni:2021fel}
R.~Casalbuoni, J.~Gomis, and D.~Hidalgo, ``{Worldline description of
  fractons},'' \href{http://dx.doi.org/10.1103/PhysRevD.104.125013}{{\em Phys.
  Rev. D} {\bfseries 104} no.~12, (2021) 125013},
  \href{http://arxiv.org/abs/2107.09010}{{\ttfamily arXiv:2107.09010
  [hep-th]}}.

\bibitem{Pena-Benitez:2021ipo}
F.~Pe\~na Benitez, ``{Fractons, Symmetric Gauge Fields and Geometry},''
  \href{http://arxiv.org/abs/2107.13884}{{\ttfamily arXiv:2107.13884
  [cond-mat.str-el]}}.

\bibitem{Angus:2021jvm}
S.~Angus, M.~Kim, and J.-H. Park, ``{Fractons, non-Riemannian geometry, and
  double field theory},''
  \href{http://dx.doi.org/10.1103/PhysRevResearch.4.033186}{{\em Phys. Rev.
  Res.} {\bfseries 4} no.~3, (2022) 033186},
  \href{http://arxiv.org/abs/2111.07947}{{\ttfamily arXiv:2111.07947
  [hep-th]}}.

\bibitem{Lake:2022ico}
E.~Lake, M.~Hermele, and T.~Senthil, ``{Dipolar Bose-Hubbard model},''
  \href{http://dx.doi.org/10.1103/PhysRevB.106.064511}{{\em Phys. Rev. B}
  {\bfseries 106} no.~6, (2022) 064511},
  \href{http://arxiv.org/abs/2201.04132}{{\ttfamily arXiv:2201.04132
  [cond-mat.quant-gas]}}.

\bibitem{Jensen:2022iww}
K.~Jensen and A.~Raz, ``{Large $N$ fractons},''
  \href{http://arxiv.org/abs/2205.01132}{{\ttfamily arXiv:2205.01132
  [hep-th]}}.

\bibitem{Brauner:2023mne}
T.~Brauner, N.~Yamamoto, and R.~Yokokura, ``{Dipole symmetries from the
  topology of the phase space and the constraints on the low-energy
  spectrum},'' \href{http://arxiv.org/abs/2303.04479}{{\ttfamily
  arXiv:2303.04479 [hep-th]}}.

\bibitem{Baig:2023yaz}
S.~A. Baig, J.~Distler, A.~Karch, A.~Raz, and H.-Y. Sun, ``{Spacetime Subsystem
  Symmetries},'' \href{http://arxiv.org/abs/2303.15590}{{\ttfamily
  arXiv:2303.15590 [hep-th]}}.

\bibitem{Kasikci:2023tvs}
O.~Kasikci, M.~Ozkan, and Y.~Pang, ``{A Carrollian Origin of Spacetime
  Subsystem Symmetry},'' \href{http://arxiv.org/abs/2304.11331}{{\ttfamily
  arXiv:2304.11331 [hep-th]}}.

\bibitem{Cheung:2023qwn}
C.~Cheung, M.~Derda, A.~Helset, and J.~Parra-Martinez, ``{Soft phonon
  theorems},'' \href{http://dx.doi.org/10.1007/JHEP08(2023)103}{{\em JHEP}
  {\bfseries 08} (2023) 103}, \href{http://arxiv.org/abs/2301.11363}{{\ttfamily
  arXiv:2301.11363 [hep-th]}}.

\bibitem{Molina-Vilaplana:2023doq}
J.~Molina-Vilaplana, ``{A post-Gaussian approach to dipole symmetries and
  interacting fractons},''
  \href{http://dx.doi.org/10.1007/JHEP08(2023)065}{{\em JHEP} {\bfseries 08}
  (2023) 065}, \href{http://arxiv.org/abs/2305.15448}{{\ttfamily
  arXiv:2305.15448 [hep-th]}}.

\bibitem{Bertolini:2023sqa}
E.~Bertolini, N.~Maggiore, and G.~Palumbo, ``{Covariant fracton gauge theory
  with boundary},'' \href{http://dx.doi.org/10.1103/PhysRevD.108.025009}{{\em
  Phys. Rev. D} {\bfseries 108} no.~2, (2023) 025009},
  \href{http://arxiv.org/abs/2306.13883}{{\ttfamily arXiv:2306.13883
  [hep-th]}}.

\bibitem{Ebisu:2023idd}
H.~Ebisu, M.~Honda, and T.~Nakanishi, ``{Foliated BF theories and Multipole
  symmetries},'' \href{http://arxiv.org/abs/2310.06701}{{\ttfamily
  arXiv:2310.06701 [cond-mat.str-el]}}.

\bibitem{Pena-Benitez:2023aat}
F.~Pe\~na Ben\'\i{}tez and P.~Salgado-Rebolledo, ``{Fracton gauge fields from
  higher dimensional gravity},''
  \href{http://arxiv.org/abs/2310.12610}{{\ttfamily arXiv:2310.12610
  [hep-th]}}.

\bibitem{Doshi:2020jso}
D.~Doshi and A.~Gromov, ``{Vortices and Fractons},''
  \href{http://arxiv.org/abs/2005.03015}{{\ttfamily arXiv:2005.03015
  [cond-mat.str-el]}}.

\bibitem{Pretko:2017kvd}
M.~Pretko and L.~Radzihovsky, ``{Fracton-Elasticity Duality},''
  \href{http://dx.doi.org/10.1103/PhysRevLett.120.195301}{{\em Phys. Rev.
  Lett.} {\bfseries 120} no.~19, (2018) 195301},
  \href{http://arxiv.org/abs/1711.11044}{{\ttfamily arXiv:1711.11044
  [cond-mat.str-el]}}.

\bibitem{Pretko:2019omh}
M.~Pretko, Z.~Zhai, and L.~Radzihovsky, ``{Crystal-to-Fracton Tensor Gauge
  Theory Dualities},''
  \href{http://dx.doi.org/10.1103/PhysRevB.100.134113}{{\em Phys. Rev. B}
  {\bfseries 100} no.~13, (2019) 134113},
  \href{http://arxiv.org/abs/1907.12577}{{\ttfamily arXiv:1907.12577
  [cond-mat.str-el]}}.

\bibitem{Gromov:2022cxa}
A.~Gromov and L.~Radzihovsky, ``{Fracton Matter},''
  \href{http://arxiv.org/abs/2211.05130}{{\ttfamily arXiv:2211.05130
  [cond-mat.str-el]}}.

\bibitem{Tsaloukidis:2023jmr}
L.~Tsaloukidis and P.~Sur\'owka, ``{Elastic Li\'enard-Wiechert potentials of
  dynamical dislocations from tensor gauge theory},''
  \href{http://arxiv.org/abs/2302.14092}{{\ttfamily arXiv:2302.14092
  [cond-mat.mtrl-sci]}}.

\bibitem{Duval:2014uoa}
C.~Duval, G.~Gibbons, P.~Horvathy, and P.~Zhang, ``{Carroll versus Newton and
  Galilei: two dual non-Einsteinian concepts of time},''
  \href{http://dx.doi.org/10.1088/0264-9381/31/8/085016}{{\em Class. Quant.
  Grav.} {\bfseries 31} (2014) 085016},
  \href{http://arxiv.org/abs/1402.0657}{{\ttfamily arXiv:1402.0657 [gr-qc]}}.

\bibitem{Bergshoeff:2014jla}
E.~Bergshoeff, J.~Gomis, and G.~Longhi, ``{Dynamics of Carroll Particles},''
  \href{http://dx.doi.org/10.1088/0264-9381/31/20/205009}{{\em Class. Quant.
  Grav.} {\bfseries 31} no.~20, (2014) 205009},
\href{http://arxiv.org/abs/1405.2264}{{\ttfamily arXiv:1405.2264 [hep-th]}}.

\bibitem{Zhang:2023jbi}
P.~M. Zhang, H.-X. Zeng, and P.~A. Horvathy, ``{MultiCarroll dynamics},''
  \href{http://arxiv.org/abs/2306.07002}{{\ttfamily arXiv:2306.07002 [gr-qc]}}.

\bibitem{Armas:2023dcz}
J.~Armas and E.~Have, ``{Carrollian fluids and spontaneous breaking of boost
  symmetry},'' \href{http://arxiv.org/abs/2308.10594}{{\ttfamily
  arXiv:2308.10594 [hep-th]}}.

\bibitem{Pretko:2018jbi}
M.~Pretko, ``{The Fracton Gauge Principle},''
  \href{http://dx.doi.org/10.1103/PhysRevB.98.115134}{{\em Phys. Rev. B}
  {\bfseries 98} no.~11, (2018) 115134},
  \href{http://arxiv.org/abs/1807.11479}{{\ttfamily arXiv:1807.11479
  [cond-mat.str-el]}}.

\bibitem{Weinberg:1965nx}
S.~Weinberg, ``{Infrared photons and gravitons},''
  \href{http://dx.doi.org/10.1103/PhysRev.140.B516}{{\em Phys. Rev.} {\bfseries
  140} (1965) B516--B524}.

\end{thebibliography}

\pagebreak

\appendix

\pagebreak
\titlepage

\setcounter{page}{1}

\begin{center}
	{\LARGE Supplemental Material}
\end{center}

\section{Soft factor from Feynman diagrams}
\label{sec:soft-factor-feynman}

In this appendix, we will derive the soft factors given
in~\eqref{eq:soft_factors} using Feynman diagram techniques. However,
when attempting to derive the soft theorems through conventional
quantum field theory methods, an immediate challenge arises. Dipoles
can be constructed from interacting monopoles are thus bound states.
Deriving an effective theory to describe its dynamics, starting from a
specific fractonic field theory, such as the one introduced by Pretko
in~\cite{Pretko:2018jbi}, is a challenging task. Fortunately, a key
attribute of the soft factor in relativistic theories is its
universality, meaning its form remains independent of the precise
theory governing the hard sector. We can naturally anticipate a
similar behavior in the case of fractons. Thus, the form of the soft
factor should be independent of the particular model used to describe
the dynamics of composite dipoles. Consequently, we will present a
simple effective model based on a worldline action, which is coupled
to the fracton gauge fields $\phi$ and $A_{ij}$ following the approach
outlined in~\cite{Figueroa-OFarrill:2023vbj}.

Let us consider the following effective model to describe the dynamics
of dipoles:
\begin{align}
	I_{\text{dip}}=\int dt\left[\frac{m}{2}\left(\frac{dz^{i}}{dt}\right)^{2}+d_{i}\frac{db^{i}}{dt}+d^{i}\partial_{i}\phi\left(t,\bm{z}\left(t\right)\right)+d^{i}\frac{dz^{j}}{dt}A_{ij}\left(t,\bm{z}\left(t\right)\right)\right] \, .
\end{align}
Here $d_{i}$ is a dipole moment degree of freedom and $b^{i}$ its
canonical conjugate. It is straightforward to show that the equation
of motion obtained from the variation of $z^{i}$ gives the Lorentz
force equation~\eqref{eq:Lorentz-force}.

In order to obtain the wave equation describing the quantum mechanics
of this model, it is useful to express the Hamiltonian action in a
form that exhibits invariance under time reparametrizations
\begin{align}
	I_{\text{dip}}=\int d\tau\left[-\pi_{t}\frac{\partial t}{\partial\tau}+\pi_{i}\frac{\partial z^{i}}{\partial\tau}-b^{i}\frac{\partial d_{i}}{\partial\tau}-N\left(\pi_{t}+d^{i}\partial_{i}\phi-\frac{1}{2m}\left(\pi_{i}-d^{j}A_{ij}\right)^{2}\right)\right] \, ,
\end{align}
where $\tau$ corresponds to the time parameter, and $N$ is a Lagrange
multiplier enforcing the constraint. To determine the wave equation
for the wave function $\psi=\psi\left(t,\bm{z},\bm{d}\right)$ we use
Dirac quantization, which means that the action of the constraint on
the physical wave functions should vanish. It is necessary to
substitute $\pi_{t}\rightarrow i\partial_{t}$ and
$\pi_{i}\rightarrow-i\partial_{i}$ which leads us to
\begin{equation}
	\label{eq:wave_eq}
	\left[i\partial_{t}+d^{i}\partial_{i}\phi-\frac{1}{2m}\left(i\partial_{i}+d^{j}A_{ij}\right)^{2}\right]\psi=0 \, .
\end{equation}
This corresponds to the Schrödinger equation and includes a coupling
to the fracton gauge fields. Note that, since the conjugate $\bm{b}$
of the dipole moment does not appear in the constraint, there are no
derivatives with respect to $\bm{d}$ in the wave equation. Therefore,
in practice, it can be treated as a fixed parameter of the problem.

The wave equation~\eqref{eq:wave_eq} can be derived from the following
field theory action
\begin{equation}
	I=\int d^{3}x\left[\psi^{*}\left(i\partial_{t}+\frac{1}{2m}\Delta\right)\psi+d^{i}\psi^{*}\psi\partial_{i}\phi+\frac{id^{j}}{2m}\left[\left(\partial_{i}\psi^{*}\right)\psi-\psi^{*}\left(\partial_{i}\psi\right)\right]A_{ij}-\frac{1}{2m}d^{j}d^{k}A_{ij}A_{ik}\psi^{*}\psi\right].\label{eq:action_dip_field}
\end{equation}
The only interacting term relevant for the emission of a single
fractonic soft photon is the one linear in $A_{ij}$. This is because
the term proportional to $\phi$ will be eliminated by the T or TL
projections, while the quartic term does not permit the emission of a
single photon.

In order to derive the soft factor one needs to introduce an external
soft photon into the hard process~\cite{Weinberg:1965nx}. This is
equivalent to adding an additional hard particle propagator and a
vertex. For the action~\eqref{eq:action_dip_field} one finds
\begin{align}
	\text{Propagator} & =\frac{i}{E-\frac{\bm{p}^{2}}{2m}} &
	\text{Vertex} & =\frac{i}{m}\epsilon_{\#}^{ij}p_{(i}d_{j)} \, .
\end{align}
If the external outgoing photon has energy $E_{\gamma}=\omega$ and
momentum $\bm{q}=\frac{\omega}{c_{\#}}\bm{n}$, and if the external
dipole has momentum $\bm{p}$ and energy $E=\bm{p}^{2}/2m$, then the
internal dipole propagator will have energy $E+\omega$ and momentum
$\bm{p}+\bm{q}$. Thus, the soft factor will be of the form
\begin{equation}
\frac{1}{\omega}\epsilon^{ij}_{\#}S_{ij}^{\#}=\lim_{\omega\rightarrow0}\frac{i}{m}\epsilon^{ij}_{\#}\left(p_{(i}+\frac{\omega}{c_{\#}}n_{(i}\right)d_{j)}\left[\frac{i}{\left(E+\omega\right)-\frac{1}{2m}\left(\bm{p}+\frac{\omega}{c_{\#}}\bm{n}\right)^{2}}\right] \, .
\end{equation}
Taking the $\omega \to 0$ limit and using that $\bm{p}=m\bm{v}$ one
finds
\begin{equation}
	\epsilon^{ij}_{\#}S_{ij}^{\#}=-\frac{\epsilon^{ij}_{\#}v_{(i}d_{j)}}{\left(1-\bm{n} \cdot \bm{v}/c_{\#}\right)} \, .
\end{equation}
Once we apply the projection induced by the polarization tensor
$\epsilon^{ij}_{\#}$ and we incorporate the $N$ external dipoles into
the process we find 
\begin{subequations}
	\label{eq:memory_dip_app}
	\begin{align}
		S_{ij}^\mathrm{T}&= - \sum_{\alpha=1}^N
		\frac{\eta^\alpha \, d^{\alpha\perp}_{(i}\,v^{\alpha\perp}_{j)}}{ 1-\bm{n} \cdot \bm{v}^\alpha /c} \label{memory T} \\
		S_{ij}^{\mathrm{TL}}&=-\sum_{\alpha=1}^N
		\frac{\eta^\alpha\,(d^{\alpha\perp}_{i}v^\alpha_{r}+v^{\alpha\perp}_{i}d^{\alpha}_r)}{ 1-\bm{n} \cdot \bm{v}^\alpha/\ct} \, . \label{memory TL}
	\end{align}
\end{subequations}
This result precisely matches the soft factors
in~\eqref{eq:soft_factors}.

\end{document}